\documentclass[aps,prl,reprint,groupedaddress,showpacs]{revtex4-1}

\usepackage[utf8]{inputenc}
\usepackage{graphicx}
\usepackage{amsmath}
\usepackage{hyperref}

\begin{document}
\title{Spherical sample holders to improve the susceptibility measurement of superparamagnetic materials}
\author{Thomas Friedrich}
\author{Tobias Lang}
\author{Ingo Rehberg}
\author{Reinhard Richter}
\affiliation{Experimentalphysik V, Universit\"at Bayreuth, D-95440 Bayreuth, Germany}
\date{\today}

\begin{abstract}
The design of two custom sample holders with a spherical cavity for commercial vibrating sample magnetometer systems is described. For such cavities the magnetization $\vec{M}$ and the internal magnetic field $\vec{H_i}$ of a sample are both homogeneous. Consequently, the material parameter $M(H_i)$ of a sample can be determined even for liquids and powders with a high magnetic susceptibility.
\end{abstract}


\maketitle

\section{Introduction}
Superparamagnetic materials like ferrofluids\,\cite{rosensweig1985} play a decisive role in modern technology. Important applications comprise vacuum tight feedthroughs in silicon crystal growing systems, the cooling of high power loudspeakers, and sink float material separation, to mention just a few\,\cite{berkovski1996}. Moreover, magnetic carriers give new perspectives in cancer therapy and other medical applications\,\cite{haefeli1997}. Consequently, there is an increasing interest in fundamental research on colloidal magnetic fluids, focusing on ferrofluid dynamics, rheology and surface instabilities\,\cite{odenbach2007}. As a result, a need arises to measure magnetic material properties with improved accuracy.

In particular, for the measurement of the magnetization of highly susceptible materials the demagnetization field has to be taken into account. It allows the determination of the magnetic field inside the sample, which is responsible for its magnetization. A homogeneous internal field can only be found in ellipsoidal samples. In the case of solids, spherical samples can be prepared by grinding\,\cite{zijlstra1967experimental}. For liquid samples and powders spherical sample holders, as introduced in this paper, seem most practical.

A very common and versatile method of measuring magnetic properties is vibrating sample magnetometry (VSM)\,\cite{Foner59}. Since its introduction, a wide range of improvements for VSM sample holders have been proposed. These include, for example, simplified handling\,\cite{drake74} or advanced temperature control\cite{defotis82,mikhov2001}. Here, we present spherical samples to determine particularly high susceptibilities with improved accuracy.

\section{Magnetometer}
A VSM determines the magnetic moment by vibrating the sample perpendicular to a uniform magnetic field in between a set of pickup coils \cite{Foner59}. Our measurements are conducted using a VSM 7404 from Lakeshore\cite{vsm7404_specs}, where the current induced in the four pickup coils is evaluated by means of a lock-in measurement. The distance between the pickup coils attached to the pole shoes is adjusted to be $25\,\mathrm{mm}$. After calibration with a reference sample, this signal provides a measure of the magnetic moment with a high sensitivity, namely a resolution of about $10^{-9}\,\mathrm{A m}^2$.

The material parameter of interest in this paper is the magnetization as a function of the internal magnetic field $M(H_i)$, but neither $M$ nor $H_i$ can be measured directly. The magnetization $\vec{M}$ must be determined from the measured magnetic moment $\vec{m}$ via
\begin{equation}
 \vec{M} = \frac{\vec{m}}{V},
\label{eq:Mvol}
\end{equation}
with $V$ being the sample volume. Consequently the measurement of the volume introduces a second source of uncertainty, which unfortunately does not match the high resolution of the magnetic measurement.

The magnetic field $\vec{H_i}$ inside a sample is in general influenced by its magnetization in a complicated way. However, in the case of an ellipsoidal sample volume, $\vec{H_i}$ can be obtained by
\begin{equation}
 \vec{H_i} = \vec{H} - \text{\textbf{D}} \vec{M}(\vec{H_i}),
\label{eq:Hint}
\end{equation}
where the external magnetic field $\vec{H}$ is measured by a Hall probe, and $\textbf{D}$ is the constant demagnetization tensor.  From a practical point of view the case of a sphere is particularly convenient, because the tensor can then be replaced by the scalar $\frac{1}{3}$.

\section{Sample Holders}
To our knowledge no commercially available sample holder has a spherical sample volume. Therefore, we have designed two spherical sample holders -- one for liquids and one for powders -- to overcome this deficiency. In the experiments presented in this paper, we compare the commercially available sample holders with the newly designed ones.

\begin{figure}
  \includegraphics[height=0.8\linewidth]{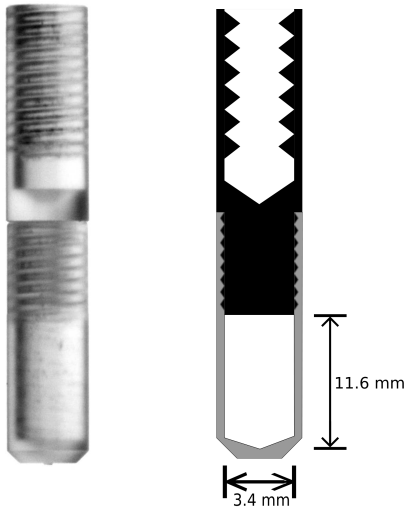}
  \caption{\label{pic:IMG_LSL} Photograph and schematic of the commercial sample holder for liquids obtained through Lakeshore (LL).}
\end{figure}

The commercially available sample holder for liquids (LL=\textbf{L}ake Shore sample holder for \textbf{L}iquids) can be obtained through Lake Shore Cryotronics Inc. (article number 730935) and is made of Kel-F\textsuperscript{\textregistered}. Figure~\ref{pic:IMG_LSL} depicts the part and shows a schematic. To get an exact measure of the sample volume, the holder is filled with a liquid of well known density (purified water at room temperature) and weighted. This way, the volume of the sample cavity can be estimated to be $69\,\mathrm{\mu l}$. In order to prevent air bubbles inside the cavity, the holder has to be overfilled. This results in an extra amount of liquid inside the thread, rendering the net volume to be $V_{\mathrm{LL}}=(80.3 \pm 9.6)\,\mathrm{\mu l}$. Consequently, the volume to calculate the magnetization of each individual sample is determined by weighting.

The cavity to contain the sample is of a complex shape with rotational symmetry. In order to calculate the internal magnetic field according to Eq.(\ref{eq:Hint}) approximately, the knowledge of an effective demagnetization tensor $\textbf{D}$ is required. Therefore, the cavity is approximated by a spheroid with the two half axes $a=1.7\,\mathrm{mm}$ and $b=5.8\,\mathrm{mm}$. According to Ref.\,\cite{osborn1945} the relevant component of the demagnetization tensor for a prolate spheroid is 
\begin{align}
 D= \frac{\frac{b}{a}}{2\left(\left(\frac{b}{a}\right)^2-1\right)} \left[ \frac{b}{a} - \frac{1}{2\sqrt{\left(\frac{b}{a}\right)^2-1}} \right. \nonumber \\
 \cdot \left. \ln{ \left( \frac{\frac{b}{a}+\sqrt{\left(\frac{b}{a}\right)^2-1}}{\frac{b}{a}-\sqrt{\left(\frac{b}{a}\right)^2-1}} \right) } \right].
\nonumber
\end{align}
In our case, this yields $D_{\mathrm{LL}}\approx0.45$.

\begin{figure}
  \includegraphics[height=0.8\linewidth,angle=0]{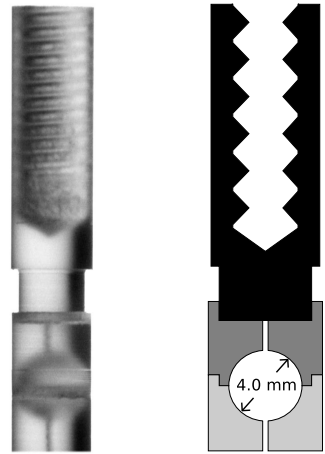}
  \caption{\label{pic:IMG_SHL} Photograph and schematic of the new sample holder for liquids (SL). The spherical cavity with a diameter of $4.0\,\mathrm{mm}$ is formed by chemical welding of two half-shells (light gray and gray). The mounting support (black) is snapped on after filling.}
\end{figure}

Figure~\ref{pic:IMG_SHL} shows the newly designed sample holder for liquids (SL=\textbf{S}pherical sample holder for \textbf{L}iquids). In order to obtain a homogeneous magnetic field inside the sample, an ellipsoidal cavity is imperative. This geometry can be achieved by chemical welding of two half shells. As it is difficult to chemically weld \mbox{Kel-F\textsuperscript{\textregistered}}, the material the commercial sample holder is made of, Makrolon\textsuperscript{\textregistered} is used instead. To overcome the volume uncertainty induced by a thread, the cavity is filled through two narrow channels with a diameter of $0.5\,\mathrm{mm}$ and a length of $1.5\,\mathrm{mm}$. After the filling process, these channels are sealed with scotch tape.

The calculated sample volume of the cavity plus the two channels is $(34.1 \pm 1.0)\,\mathrm{\mu l}$. The given error accounts for engineering tolerances. To measure the sample volume, the holder is filled with water and weighted. By this, the sample volume results to be $V_{\mathrm{SL}}=(35.3 \pm 0.2)\,\mathrm{\mu l}$, which is consistent with the calculation. As with the LL, the magnetization is calculated using the individual sample volume determined by weighting.

\begin{figure}
  \includegraphics[height=0.8\linewidth]{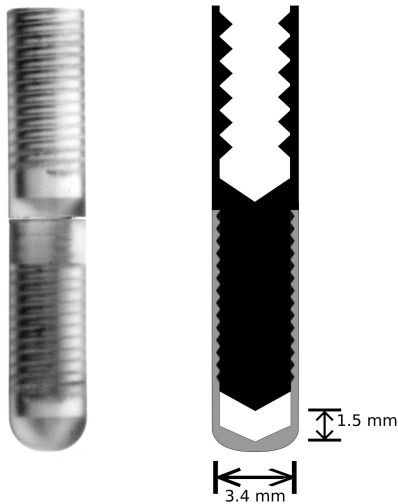}
  \caption{\label{pic:IMG_LSP} Photograph and schematic of the commercial sample holder for powders obtained through Lakeshore (LP).}
\end{figure}

The commercially available sample holder for powders (LP=\textbf{L}ake Shore sample holder for \textbf{P}owders) obtained from Lake Shore Cryotronics Inc. (article number 730931) is made of Kel-F\textsuperscript{\textregistered}. It is depicted in Fig.\,\ref{pic:IMG_LSP}.

In this case, filling the sample holder with water and weighting gives only a rough estimate of the volume. This is partly due to the relatively small volume and partly due to the nuisance, that the water tends to creep up the thread. The measured volume is $V_{\mathrm{LP}} = (15 \pm 5)\,\mathrm{\mu l}$. Weighting of a sample does not yield its gross volume, because that depends on its packing density. Therefore, the volume of the sample cavity $V_{\mathrm{LP}}$ is used to calculate the magnetization.

As indicated by the sketch, the shape of the sample chamber is again different from an ellipsoid. Similar to the LL, the cavity is approximated by an oblate spheroid with the two half axes $a=1.7\,\mathrm{mm}$ and $b=0.75\,\mathrm{mm}$. According to Ref.\,\cite{osborn1945} the relevant component of the demagnetization tensor is 
\begin{align}
 D=\frac{1}{2\left(\left(\frac{a}{b}\right)^2-1\right)} \left[ \frac{\left(\frac{a}{b}\right)^2}{\sqrt{\left(\frac{a}{b}\right)^2-1}} \right. \nonumber \\
 \left. \cdot \arcsin{\left(\sqrt{1-\left(\frac{b}{a}\right)^2} \right) -1} \right].
\nonumber
\end{align}
In our case, this yields $D_{\mathrm{LP}}\approx 0.22$.

\begin{figure}
  \includegraphics[height=0.8\linewidth,angle=0]{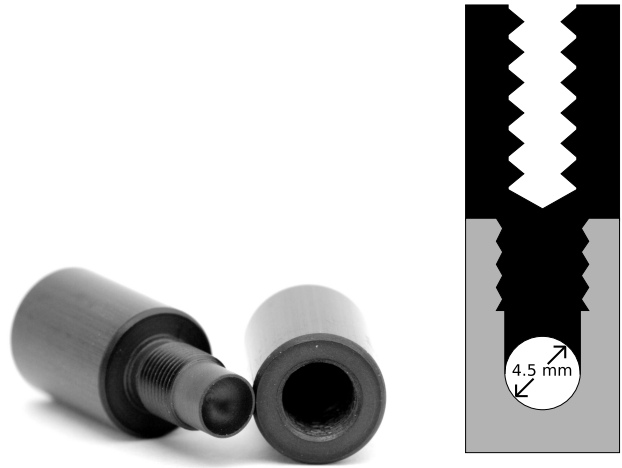}
  \caption{\label{pic:IMG_SHP} Photograph and schematic of the new sample holder for powders (SP). The spherical cavity formed by the two half-shells has a diameter of $4.5\,\mathrm{mm}$.}
\end{figure}

Since the SL shown in Fig.\,\ref{pic:IMG_SHL} cannot be filled with powders, we designed the sample holder (SP=\textbf{S}pherical sample holder for \textbf{P}owders) shown in Fig.\,\ref{pic:IMG_SHP}. A well suited material, with special respect to the sharp edges at the bottom of the upper part, is Vespel\textsuperscript{\textregistered}. Taking engineering tolerances into account, the volume of the cavity can be calculated to $V_{\mathrm{SP}}=(47.7 \pm 1.3)\,\mathrm{\mu l}$. An independent measurement by the water weighting method used for the other three sample holders is not possible here due to the special geometry. As with the LP, the volume of the sample cavity $V_{\mathrm{SP}}$ has to be used to calculate the magnetization.

\section{Experimental Results and Discussion}
To compare the sample holders, three different substances -- water, ferrofluid and magnetite powder -- are used. 

The water is purified (SG LaboStar 7 TWF-DI). It qualifies as an adequate reference fluid because its composition and susceptibility are very well defined. The magnetic susceptibility is known to be \mbox{$\chi_{\mathrm{water}}=-9.0238\cdot 10^{-6}$}\,\cite{CRC2009}.   

The ferrofluid (APG512a from FerroTec Corp.) is based on synthetic esters and contains magnetite nano particles. It is a good testing fluid due to its long-term stability (years)\,\cite{gollwitzer2009}. Because of its high susceptibility $\chi \approx 1$ it is suitable to demonstrate the impact of the different geometries.

For the powder sample holders, magnetite ($\mathrm{Fe}_3\mathrm{O}_4$) has been obtained through Sigma-Aldrich Co., consisting of spherical nano particles with a size of \mbox{$d < 50\,\mathrm{nm}$}\,\cite{Takacs07,Peng09}. The powder represents a different class of materials with specific intricacies. 

The measurement procedure consists of i) setting the external field to $800\,\mathrm{kA} / \mathrm{m}$, ii) decreasing it step by step to $-800\,\mathrm{kA} / \mathrm{m}$, and iii) increasing it back to $800\,\mathrm{kA} / \mathrm{m}$. In each step, the component $m_{\parallel}$ of the magnetic moment $\vec{m}$ parallel to the external field $\vec{H}$ is recorded by the VSM.

\subsection{Water}
The determination of the magnetization consists of two independent measurements. First, the magnetic moment of the sample as a function of the external magnetic field is measured with a high accuracy by the VSM. Second, the volume of the sample is determined as described above. These two quantities yield the volume magnetization $M$ as stated in Eq.\,(\ref{eq:Mvol}). In order to validate this two-step determination, water is used as a reference fluid with a well known constant susceptibility. In addition, its demagnetization field can be neglected due to the small susceptibility. Therefore one can expect, that both sample holders produce the same result. Any deviations must be attributed to the different volumes, geometries and the different spatial distribution of the liquid.

\begin{figure}
  \includegraphics[width=1.0\linewidth]{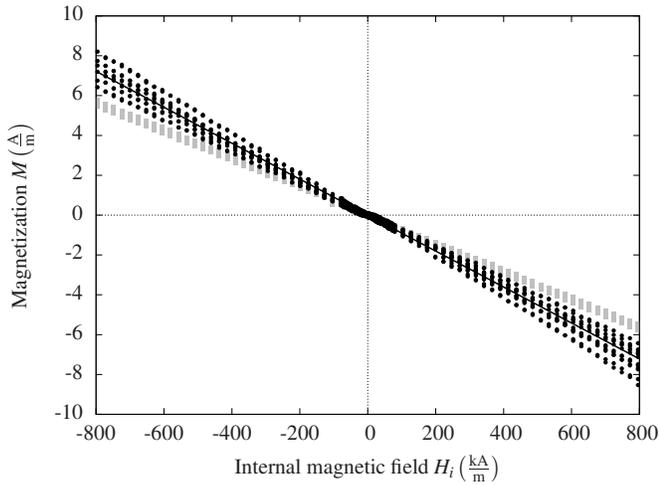}
  \caption{\label{plot:water_MvsH} Comparison of the measured magnetization curves of purified water. The magnetization per volume is plotted as a function of the internal magnetic field for the commercially available LL (gray dots) and the SL (black dots). The solid black line marks the literature value\cite{CRC2009}.}
\end{figure}

The results for water are depicted in Fig.\,\ref{plot:water_MvsH}. Here, the magnetization is plotted as a function of the internal magnetic field. The magnitude of the internal field is deduced via
\begin{equation}
  H_i = H - D M(H_i).
\end{equation}
Note, that the correction due to the second term is only about 4\,ppm for the case of water, but was used nontheless for the sake of a standardized data processing.
The external magnetic field $H$ is measured by the Hall probe in a distance of $9\,\mathrm{mm}$ from the sample\,\cite{vsm7404_specs}. As explained above, $D=D_{\mathrm{LL}}$ was used for the LL and \mbox{$D=\frac{1}{3}$} for the SL. The magnetization $M(H_i)$ is calculated via 
\begin{equation}
 M=\frac{m_{\parallel}-m_{\parallel,0}}{V}.
\end{equation}
Here the volume $V$ is determined by weighting of the sample. The magnetic moment of the empty sample holder $m_{\parallel,0}$ and of the measured water sample $m_{\parallel}$ are of the same order of magnitude, therefore the signal from the empty sample holder has been subtracted.

\begin{figure}
  \includegraphics[width=1.0\linewidth]{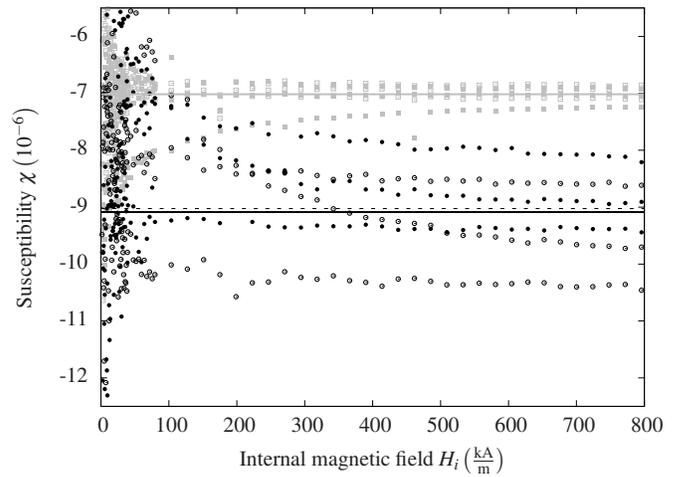}
  \caption{\label{plot:water_chivsH} Comparison of the measured susceptibilities of purified water. The susceptibility per volume is plotted as a function of the internal magnetic field for the commercially available LL (gray dots) and the SL (black dots). The solid gray line denotes the mean value of the gray points, and the solid black line shows the mean of the black points. The dashed black line marks the literature value\cite{CRC2009}.}
\end{figure}

In Fig.\,\ref{plot:water_MvsH} the results for the LL are shown in gray and those for the SL in black. In order to emphasize the differences between both measurements, these data are replotted in Fig.\,\ref{plot:water_chivsH}, which depicts the susceptibility \mbox{$\chi=M / H_i$} as a function of the internal magnetic field, where $M$ is an average according to
\begin{equation}
M=\frac{M_{\mathrm{up},H > 0}+M_{\mathrm{down},H>0}-M_{\mathrm{up},H<0}-M_{\mathrm{down},H<0}}{4}.
\label{eq:M_avg}
\end{equation}

The susceptibilities obtained with the LL show a smaller scattering. This is partly due to the larger sample volume of the LL and partly due to the smaller magnetic moment of the empty sample holder. More importantly, the mean values of the susceptibilities shown as solid lines differ significantly. The values are listed in Table \ref{table:chi_h2o}, the given errors reflect the largest deviation of an individual measurement from the mean value. Although the susceptibilities from the SL show a higher scattering, their average $\chi_{\mathrm{SL}}$ reproduces the literature value $\chi_{\mathrm{water}}$ quite well. On the other hand, the comparison of $\chi_{\mathrm{LL}}$ to $\chi_{\mathrm{water}}$ shows a systematic deviation of $29\%$. This is believed to be caused by the sensitivity function of the pickup system, as stated in the VSM systems documentation\,\cite{vsm7404_doc}. For a sample larger than the Ni-sphere with a diameter of $2\,$mm, which is provided by LakeShore as a reference sample, this effect gains momentum. To quantify this deviation, we introduce a correction factor $Q_\mathrm{holder}=\frac{\chi_{\mathrm{water}}}{\chi_\mathrm{holder}}$. According to Table\,\ref{table:chi_h2o}, we determine $Q_\mathrm{LL}=(1.29 \pm 0.1)$ and $Q_\mathrm{SL}=(0.99 \pm 0.03)$. $Q_{\mathrm{SL}}$ cannot be smaller than unity. We thus replace it by $Q_{\mathrm{SL}}=1$, and conclude that the deviation from unity is beyond our resolution.
\begin{table}
\centering
\begin{tabular}{c|l}
 $\chi_{\mathrm{LL}}$		&	$(-7.017 \pm 0.3)\cdot 10^{-6}$	\\ \hline
 $\chi_{\mathrm{SL}}$		&	$(-9.086 \pm 1)\cdot 10^{-6}$	\\ \hline
 $\chi_{\mathrm{water}}$	&	$-9.0238\cdot 10^{-6}$	
\end{tabular}
\caption{Extracted susceptibilities for water and a literature value.}
\label{table:chi_h2o}
\end{table}

\subsection{Ferrofluid}
\begin{figure}
  \includegraphics[width=1.0\linewidth]{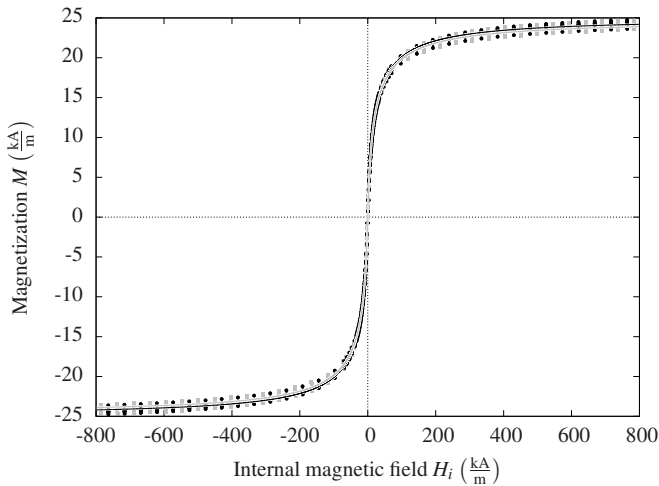}
  \caption{Comparison of the measured magnetization curves of the ferrofluid APG512a. The magnetization per volume is plotted as a function of the internal magnetic field for the commercially available LL (gray) and the SL (black). For clarity, only every second data point is shown. The solid lines indicate fits to the MMF2 model.}
\label{plot:apg_MvsH}
\end{figure}

The results for APG512a are depicted in Fig.\,\ref{plot:apg_MvsH}. Here, the magnetization is plotted as a function of the internal magnetic field. We use the same data processing as for water, except for the subtraction of the magnetic moment of the empty sample holder. This is justified, because the magnetic moment of the ferrofluid is three orders of magnitude higher than that of the sample holder. The magnetization $M=\frac{m_{\parallel}}{V} \cdot Q_{\mathrm{holder}}$ for the LL is shown in gray and the one for the SL in black. The scattering with the SL is of similar magnitude compared to the LL even though the sample volume of the SL is smaller. The data can be approximated by the so called second order modified mean-field model for dense ferrofluids (MMF2) described in Refs.\,\cite{ivanov2001,ivanov2007a}. Nonlinear curve fitting by this model yields the saturation magnetization and the initial susceptibility. The extracted values are shown in Tab.\,\ref{table:apg512a}. The given error for the saturation magnetization reflects the largest deviation of an individual measurement from the mean value. The error for the susceptibility is estimated by a visual inspection of the slope in the vicinity of the origin. The comparison of the measured saturation magnetization for the SL with the value given in the data sheet shows a good agreement within the given tolerances.

\begin{figure}
  \includegraphics[width=1.0\linewidth]{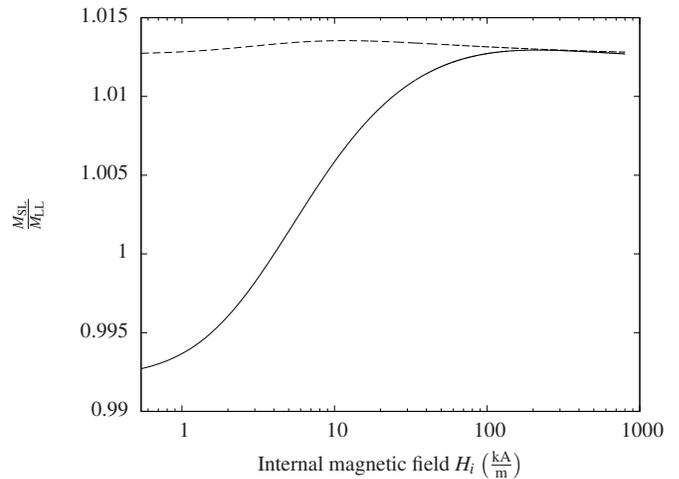}
  \caption{The ratio of the magnetizations obtained by the two sample holders for APG512a. The solid line is based on the demagnetization factor $D_{\mathrm{LL}}$. The dashed line is obtained by using the empirical demagnetization factor $D_{\mathrm{LL,emp}}$.}
\label{plot:apg_quotient}
\end{figure}

In order to accentuate the differences between the two magnetization curves, the ratio of the fitted model functions is plotted in Fig.\,\ref{plot:apg_quotient}. The solid line shows that the saturation magnetization measured with the LL is by $1\%$ too small, compared to the SL. For weak magnetic fields, the deviation between the two sample holders becomes more important. The initial susceptibilities differ by about $1\%$ in the opposite direction. This must be attributed to the demagnetization factor because it shows its highest influence in this regime. We conclude, that the measurements utilizing a spherical cavity are superior in this regime because the irregular shaped sample holder does not have a well defined demagnetization factor. It seems worth pointing out, that this deficiency can partly be corrected by means of an empirical demagnetization factor of $D_{\mathrm{LL,emp}}=0.438$. This value has been obtained under the condition that the ratio of the initial susceptibilities matches the ratio of the saturation magnetizations. Note that this corrected ratio (dashed line) still changes as a function of the magnetic field, which is due to the fact that a constant demagnetization factor does not exist for a non-elliptical geometry.
\begin{table}
\centering
\begin{tabular}{c|c|c}
		&	$M_{\mathrm{sat}}(\frac{\mathrm{kA}}{\mathrm{m}})$		&	$\chi_0$	\\ \hline
 LL		&	$24.74 \pm 0.5$	&	$1.404 \pm 0.06$	\\ \hline
 SL		&	$25.05 \pm 0.4$	&	$1.393 \pm 0.06$	\\ \hline
 Data sheet	&	$26 \pm 2.6$	&	n.a.	
\end{tabular}
\caption{Extracted saturation magnetizations and initial susceptibilities for APG512a.}
\label{table:apg512a}
\end{table}

\subsection{Powder}
\begin{figure}
  \includegraphics[width=1.0\linewidth]{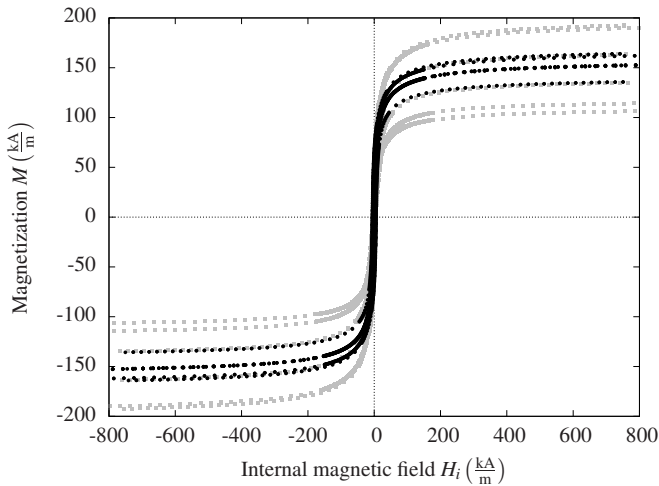}
  \caption{Comparison of the measurements of the magnetite powder. The magnetization is plotted as a function of the internal magnetic field for the commercially available LP (gray) and the SP (black).}
\label{plot:Magnetite_MvsH}
\end{figure}

We characterize the magnetic properties of the powder by $M=\frac{m_{\parallel}}{V}$ where $V$ is either $V_\mathrm{LP}$ or $V_\mathrm{SP}$. The magnetization as a function of the internal field for the LP is calculated using the approximated demagnetization factor $D_{\mathrm{LP}}$. To compare both holders the magnetization curves for magnetite powder are depicted in Fig.\,\ref{plot:Magnetite_MvsH}. While the average magnetization for both sample holders agrees fairly well, the scattering for the LP is significantly higher than for the SP. This is believed to be caused by the high uncertainty of the sample volume in this case, which is much harder to control than for fluids. On the one hand, filling the sample volume with excess material (as done for fluids) is prohibitive because it cannot be squeezed out of the measuring volume. On the other hand, a lack of material cannot as easily be detected as an air bubble in the case of a fluid. Concerning both aspects, the new sample holder is superior because a lack of material can easily be seen by reopening the sample holder (a defect becomes visible in the spherical surface of the pressed powder), and an excess prevents a closing of the thread. 

\begin{figure}
  \includegraphics[width=1.0\linewidth]{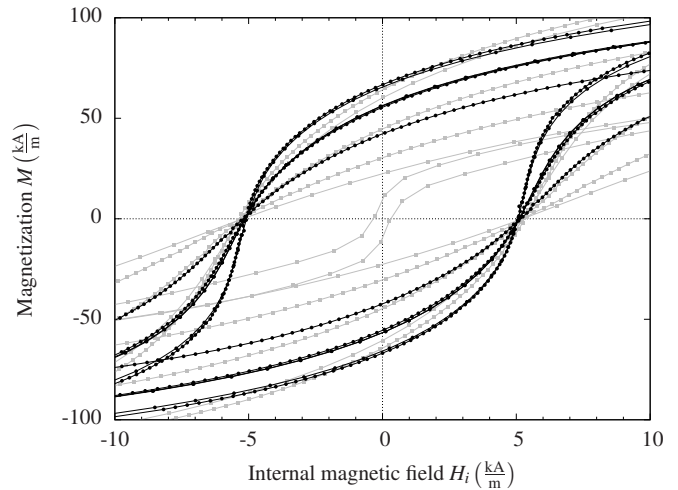}
  \caption{Magnification of the central region of Fig.\,\ref{plot:Magnetite_MvsH}. The magnetization is plotted as a function of the internal magnetic field for the commercially available LP (gray) and the SP (black). The points are connected by lines to guide the eyes.}
\label{plot:Magnetite_MvsH_hysterese}
\end{figure}

Figure~\ref{plot:Magnetite_MvsH_hysterese} magnifies the central region of Fig.\,\ref{plot:Magnetite_MvsH} to unveil the hysteretic behavior of the magnetite powder. The dataset with the smallest saturation magnetization shows only a very tiny hysteresis. We cannot explain this behaviour, but suspect it to be caused by an underfilling of the cavity, allowing a part of the material to move freely within the sample holder. The other curves show a clearly detectable hysteresis with the expected behavior, that they all show the same coercivity. This reflects the fact, that neither the demagnetization factor nor the sample volume have an influence in the case of vanishing magnetization.

\begin{figure}
  \includegraphics[width=1.0\linewidth]{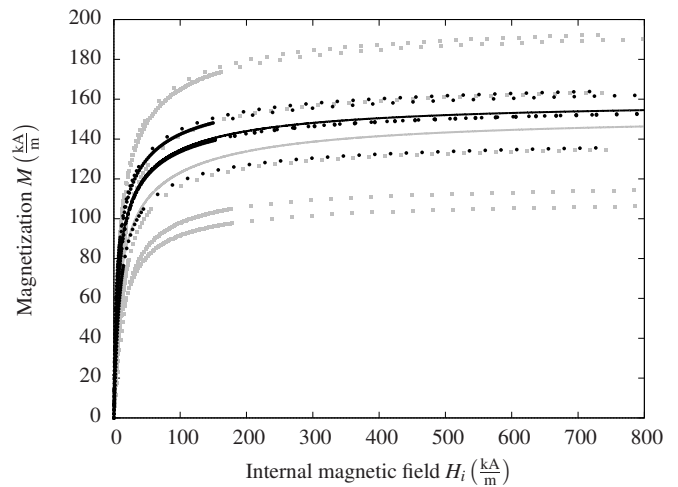}
  \caption{Averaged magnetization curves according to Eq.\,(\ref{eq:M_avg}) for magnetite powder measured with the SP (black) and the LP (gray). The solid lines indicate fits to the MMF2 approximation.}
\label{plot:Magnetite_kph_MvsH}
\end{figure}

In order to extrapolate the saturation magnetization and the initial susceptibility, we use a similar procedure as for the ferrofluid. In this case, however, prior to the fitting we calculate an average magnetization according to Eq.\,(\ref{eq:M_avg}). Note, that this procedure is only used as a convenient method for extrapolating the saturation magnetization and the initial susceptibility. The corresponding results are shown in Fig.\,\ref{plot:Magnetite_kph_MvsH}. We can successfully fit them by the MMF2, as indicated by the solid lines, although the underlying theory is designed for paramagnetic behaviour. The outcome is summarized in Tab.\,\ref{table:magnetit}. The errors are deduced as described for Tab.\,\ref{table:apg512a}.
\begin{table}
\centering
\begin{tabular}{c|c|c}
		&	$M_{\mathrm{sat}}(\frac{\mathrm{kA}}{\mathrm{m}})$		&	$\chi_0$	\\ \hline
 LP		&	$153 \pm 43$	&	$17.760 \pm 3$	\\ \hline
 SP		&	$159 \pm 14$	&	$18.461 \pm 1$	\\ \hline
 Data sheet	&	$446$			&	n.a.	
\end{tabular}
\caption{Extracted saturation magnetizations and initial susceptibilities for magnetite powder.}
\label{table:magnetit}
\end{table}

The average saturation magnetization \mbox{$M_{\mathrm{sat,exp}} = 156\,\mathrm{kA/m}$} has to be compared to $M_{\mathrm{sat,lit}}=446\,\mathrm{kA} / \mathrm{m}$ \cite{rosensweig1985}. The reduction to $35\%$ can partly be attributed to the filling fraction of the powder of about $42\%$, and partly to an oxidation of the surface of the individual grains.

\begin{figure}
  \includegraphics[width=1.0\linewidth]{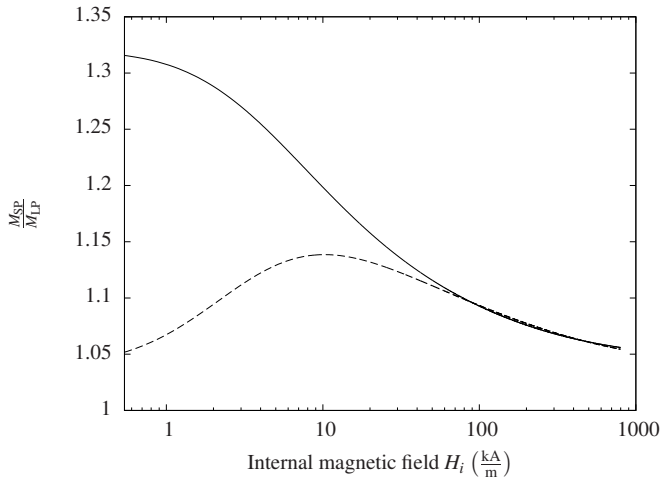}
  \caption{The ratio of the magnetizations obtained by the two sample holders. The solid line is based on the demagnetization factor $D_{\mathrm{LP}}$. The dashed line is obtained by using the empirical demagnetization factor $D_{\mathrm{LP,emp}}$.}
\label{plot:Magnetite_quotient}
\end{figure}

The ratio of the fitted model functions is plotted in Fig.\,\ref{plot:Magnetite_quotient}. The solid line shows that the saturation magnetization measured with the LP is by $4\%$ too small, compared to the SP. As for the ferrofluid, the deviation becomes more important for weak magnetic fields. For the presented data, this deficiency can partly be corrected by means of an empirical demagnetization factor of $D_{\mathrm{LP,emp}}=0.236$. In contrast to the LL, that correction is less satisfactory in this case leaving a $10\%$ deviation for an internal magnetic field of $10\,\mathrm{kA} / \mathrm{m}$. This reflects the fact that an ellipsoid is a better approximation for the LL compared to the LP.

\section{Conclusion}
In this paper we have introduced two sample holders for VSM measurements. They have a well defined demagnetization factor which makes them suitable for the investigation of superparamagnetic materials. In addition to being superior in principle, these sample holders even provide the advantage of a better defined filling procedure in particular for magnetic powders. Moreover, our investigation leads to empirical demagnetization factors for the commercially available sample holders.

\begin{acknowledgments}
The authors are grateful to DFG FOR608 for financial support. Further, we thank K. Oetter for fructuous discussion and for the prototyping of the sample holders.
\end{acknowledgments}

\providecommand{\noopsort}[1]{}\providecommand{\singleletter}[1]{#1}%

\end{document}